\documentclass[conference]{IEEEtran}
\IEEEoverridecommandlockouts
\usepackage[utf8]{inputenc}
\usepackage[T1]{fontenc}
\usepackage{amsmath,amssymb,bm}
\usepackage{graphicx}
\usepackage{booktabs}
\usepackage{array}
\usepackage{xcolor}
\usepackage[hidelinks]{hyperref}

\begin{document}

\title{Moving Horizon Estimation and Constrained Model Predictive Control for Balancing a Wheel-Legged Robot Under Unknown Disturbances}

\author{\IEEEauthorblockN{Haitham El-Hussieny}
	\IEEEauthorblockA{\textit{Department of Mechatronics and Robotics Engineering}\\\textit{Egypt-Japan University of Science and Technology (E-JUST)}\\
		Alexandria, Egypt\\
		haitham.elhussieny@ejust.edu.eg}}
\maketitle

\begin{abstract}
Wheel-legged robots balance on two wheels while using articulated legs to change height, absorb impacts, and cross uneven terrain. On hardware, the balance controller must rely on noisy inertial and encoder measurements, and it must reject disturbances that are not known in advance, such as payloads and wind, without exceeding motor torque limits. This paper presents a moving horizon estimator (MHE) combined with an offset-free, constrained model predictive controller (MPC) for the sagittal balance of a 9.95 kg wheel-legged robot. The estimator uses a virtual-leg model augmented with a horizontal hip force, a body pitch torque, and a gyroscope bias. Its accelerometer model removes the predicted hip acceleration, and its process noise is increased when an impact is detected. We evaluate the approach in MuJoCo with simulated sensor noise, bias, quantization, and delay, and compare it with LQR baselines that use ground-truth state, a complementary filter, or the same MHE. In a driving, squatting, reversing, and bowing sequence, MPC with MHE reduced the RMS speed error to 0.11 m/s, compared with 0.24 m/s for LQR with ground-truth state. Under an unknown 2 kg payload and a 10 N headwind, the mean position error in wind was 6 cm, compared with 81 cm for LQR with ground-truth state. Across a 3 $\times$ 4 grid of payloads and headwinds, LQR with the same estimator fell in 3 of 12 combinations, compared with 1 of 12 for MPC. These results are limited to simulation on flat ground, and experimental validation on hardware remains future work.
\end{abstract}


\section{Introduction}
\emph{Wheel-legged robots} combine the efficiency of wheeled locomotion with the terrain adaptability of legs \cite{klemm2020}. Such a robot rolls on its wheels for most of its motion, and its legs change the height of the body, absorb impacts, and step over obstacles, making them well suited to exploring challenging environments, where terrain cannot be mapped in advance \cite{elhussieny2013}. Balance is the central control problem \cite{zhang2025balance}. The body is an inverted pendulum on wheels, so it must be stabilized continuously, and the stabilizing torque must come through the same wheel motors that drive the robot \cite{liu2024diablo}.

Most balance controllers assume that the full state is measured. On hardware this assumption fails. The inertial measurement unit (IMU) reports body rates with bias and noise, the body pitch must be inferred from an accelerometer that also responds to the robot's own acceleration, and wheel encoders give velocity only to finite resolution. Errors in the state estimate enter the balance loop directly. A disturbance that the controller was not told about, such as a payload or a gust of wind, produces a steady drift. The controller must correct that drift with torque, and the torque bound limits how much correction is available.

Estimation and control are usually designed separately. A complementary filter or Kalman filter feeds a linear-quadratic regulator (LQR). This separation ignores two couplings. First, the filter does not know which disturbances the controller must reject. Second, the regulator does not know that its inputs are bounded. A complementary filter is fast and simple, but it assumes that sensor errors are small and slowly varying \cite{mahony2008}, and it does not model the disturbance it must reject. An LQR gain is optimal for the linearized, unconstrained problem \cite{rawlings2017}, so its guarantees weaken once torques saturate.

This paper combines the two couplings in one design. The contributions are:
\begin{itemize}
\item a moving horizon estimator (MHE) whose model includes the unknown hip force, the body pitch torque, and the gyroscope bias as disturbance states, and whose accelerometer measurement removes the predicted hip acceleration;
\item an impact-aware process-noise rule that prevents a ground impact from being interpreted as a change in the robot state;
\item an offset-free constrained MPC that uses the estimated disturbances, together with a comparison against LQR baselines that share the same sensor simulation and the same discrete gain.
\end{itemize}
We do not claim a new balance law. The novelty lies in the combination and in a controlled comparison under realistic sensing. All results are from simulation, and hardware validation remains future work.

The remainder of this paper is organized as follows. Section \ref{sec:rw} reviews balance estimation, control of wheel-legged robots, and model predictive control. Section \ref{sec:model} presents the robot model and its virtual-leg linearization. Section \ref{sec:mhe} describes the moving horizon estimator, including the augmented disturbance model, the accelerometer measurement, and the impact handling. Section \ref{sec:mpc} presents the LQR baselines and the offset-free constrained MPC. Section \ref{sec:setup} describes the simulation setup, the sensor model, and the three scenarios. Section \ref{sec:results} reports the results and the robustness grid. Section \ref{sec:discussion} discusses the findings and their limitations, and Section \ref{sec:conc} concludes the paper.

\section{Related Work}
\label{sec:rw}
\textbf{Balance estimation.} Three families of estimator are common. Complementary and Mahony-type filters are inexpensive and robust to sensor drift, but they do not use the robot model \cite{mahony2008}. Kalman filters use the model and noise statistics, and they are optimal for linear Gaussian problems \cite{kalman1960}. Moving horizon estimation solves a constrained least-squares problem over a sliding window of measurements. It handles bounds and nonlinear models, at a higher computational cost \cite{rao2003, you2021state}.

\textbf{Control of wheel-legged robots.} One approach is linear optimal control with gain scheduling over leg length. It is simple, fast, and well understood, but it cannot represent input bounds and it is tuned for a single operating region \cite{klemm2020}. Another approach is whole-body or hierarchical control, which coordinates the wheels and legs through task priorities. It handles more contacts but requires more modeling \cite{wen2025wbc}. A third approach is learned control, which can adapt to the robot but requires extensive training data and is difficult to certify \cite{elhussieny2024real}.

\textbf{Model predictive control.} MPC solves a finite-horizon optimal control problem at every sample and applies its first input. It handles state and input constraints explicitly \cite{rawlings2017}. Its cost is a per-sample optimization, which must fit within the control period. Offset-free MPC adds a disturbance model whose estimate removes steady-state error caused by unmeasured inputs \cite{pannocchia2003}. The present work combines an MHE that provides the disturbance estimate with such an offset-free MPC.

\section{Robot Model}
\label{sec:model}
\begin{figure}[t]
\centering
\includegraphics[width=.9\columnwidth]{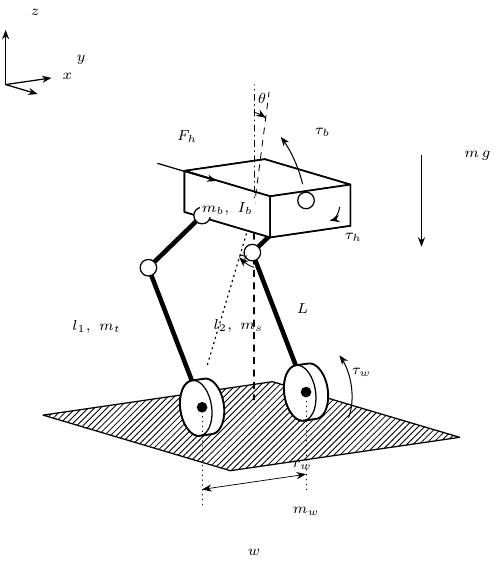}
\caption{Three-dimensional sketch of the robot with nominal dimensions and masses. The hip $H$ is the body origin, the knee $K$ joins thigh and shank, and the wheel axle $W$ is at the end of the virtual leg. The virtual leg of length $L$ lumps the two-link leg and lies in the sagittal plane, with angle $\gamma$ from the vertical. The body pitch $\theta$ is measured from the vertical. The forces and torques are the hip force $F_h$, the body pitch torque $\tau_b$, the hip torque $\tau_h$, and the wheel torque $\tau_w$. Values are listed in Table~\ref{tab:params}. The knee torque is set by the leg-length controller and is not an input of the reduced model.}
\label{fig:robot}
\end{figure}

The robot, shown in Fig.~\ref{fig:robot}, has a mass of 9.95 kg, two drive wheels, and two legs. Its dimensions and masses are listed in Table~\ref{tab:params}. Each leg is represented as a virtual leg: a massless prismatic element of length $L$ between the hip and the wheel axle. The body is a rigid mass with inertia. The sagittal motion is described by the generalized coordinates $\bm q = [x,\ \gamma,\ \theta]^\top$, where $x$ is the wheel-axle position, $\gamma$ is the leg angle from the vertical, and $\theta$ is the body pitch. The inputs are $\bm u = [\tau_w,\ \tau_h]^\top$, the wheel torque and the hip torque.

\begin{table}[t]
\centering
\caption{Robot dimensions, masses, and actuator limits from the MuJoCo model. Masses marked per leg or per wheel are for each of the two legs or wheels.}
\label{tab:params}
\begin{tabular}{@{}llr@{}}
\toprule
Symbol & Quantity & Value \\
\midrule
$m_b$ & body mass & 8.27 kg \\
$m_t$ & thigh mass, per leg & 0.25 kg \\
$m_s$ & shank mass, per leg & 0.29 kg \\
$m_w$ & wheel mass, per wheel & 0.30 kg \\
$m$ & total mass, $m_b+2(m_t+m_s+m_w)$ & 9.95 kg \\
$I_b$ & body pitch inertia & 0.094 kg\,m$^2$ \\
$l_1$ & thigh length & 0.250 m \\
$l_2$ & shank length & 0.337 m \\
$r_w$ & wheel radius & 0.0599 m \\
$w$ & track width & 0.272 m \\
$L$ & virtual leg length (operating range) & 0.22--0.56 m \\
$g$ & gravitational acceleration & 9.81 m\,s$^{-2}$ \\
$\tau_h$ & hip torque limit & $\pm 45$ N\,m \\
$\tau_w$ & wheel torque limit, per wheel & $\pm 10$ N\,m \\
$\bm u_{\max}$ & controller input bounds (total wheel, hip) & $(16,\ 40)$ N\,m \\
\bottomrule
\end{tabular}
\end{table}

Leg force is computed by a stiffness-and-damping law, $F_L = F_g + k_L(L_d - L) - d_L\dot L$, and mapped to the joints through the leg Jacobian. This regulates leg length. The dynamics are linearized about the upright posture for each leg length in a grid of 12 points spanning 0.22--0.56 m. A disturbance vector $\bm d = [F_h,\ \tau_b]^\top$ collects the horizontal hip force and the body pitch torque. The linearized model is
\begin{equation}
\dot{\bm x} = \bm A(L)\,\bm x + \bm B(L)\,\bm u + \bm E(L)\,\bm d,
\label{eq:lin}
\end{equation}
where $\bm x = [\bm q^\top,\ \dot{\bm q}^\top]^\top \in \mathbb{R}^6$, $\bm A(L)$ and $\bm B(L)$ are the state and input matrices, and $\bm E(L)$ maps the disturbance into the dynamics. The LQR gain is scheduled on $L$ by interpolation over the grid. The full Lagrangian formulation and leg-force gains are given in the accompanying code repository.

\section{Moving Horizon Estimator}
\label{sec:mhe}
\subsection{Augmented model}
The estimator state is $\bm z = [\bm x^\top,\ \bm d^\top,\ b]^\top \in \mathbb{R}^9$, where $b$ is the pitch-gyroscope bias. The disturbance and bias are modeled as random walks, so that they are estimated rather than assumed. Discretizing Eq.~\eqref{eq:lin} at $T_s = 10$ ms with a zero-order hold gives
\begin{equation}
\bm z_{k+1} = \bm A_z(L_k)\,\bm z_k + \bm B_z(L_k)\,\bm u_k + \bm w_k,
\label{eq:aug}
\end{equation}
where $\bm w_k$ is process noise with covariance $\bm Q_k$. The model matrices are precomputed on an 18-point leg-length grid.

\subsection{Measurements}
The measurement vector combines four sources. Wheel odometry provides position and velocity. Joint encoders provide the leg-to-body angle $\gamma - \theta$. The gyroscope provides the pitch rate $\dot\theta + b$. The accelerometer provides the pitch information that is needed when odometry drifts. The accelerometer measures the specific force, which includes the robot's own horizontal hip acceleration. The model predicts that acceleration, so the measurement model subtracts it:
\begin{equation}
y_{\mathrm{acc}} = \theta - \frac{(\bm A_3 + L\,\bm A_4)\,\bm x}{g} - \frac{(\bm E_3 + L\,\bm E_4)\,\bm d}{g} + v_{\mathrm{acc}},
\label{eq:acc}
\end{equation}
where $g$ is gravitational acceleration, $\bm A_i$ and $\bm E_i$ are the $i$-th rows of $\bm A$ and $\bm E$ (for the pitch and horizontal-acceleration equations), and $v_{\mathrm{acc}}$ is measurement noise. Without this correction, a strong acceleration during driving would be read as a change in body pitch. The correction is applied only when the accelerometer is trusted (see below).

\subsection{Optimization problem}
At each sample the MHE solves a bounded least-squares problem over the last $N=25$ samples, which is a 250 ms window at 100 Hz:
\begin{align}
\min_{\bm z_{0:N}, \bm w_{0:N-1}, \bm v_{0:N}}\ &\|\bm z_0 - \bar{\bm z}\|^2_{\bm P^{-1}} + \sum_{k=0}^{N-1}\|\bm w_k\|^2_{\bm Q^{-1}}  \nonumber\\
&+ \sum_{k=0}^{N}\|\bm v_k\|^2_{\bm R^{-1}} \nonumber\\
\text{s.t.}\ & \text{Eq.~\eqref{eq:aug} and Eq.~\eqref{eq:acc}}, \nonumber\\
& |\hat d_i| \le d_{\max,i},\quad |\hat b| \le b_{\max},
\label{eq:mhe}
\end{align}
where $\bar{\bm z}$ is the arrival state and $\bm P$ is the arrival covariance, obtained from the discrete algebraic Riccati equation. The disturbance bounds are $d_{\max} = (60\,\mathrm{N},\ 15\,\mathrm{N\,m})$ and the bias bound is $b_{\max} = 0.15$ rad/s. The problem is solved with OSQP \cite{stellato2020} with polishing disabled. The mean solve time was 3.5 ms in scenario A, with a maximum of 17.9 ms (Section~\ref{sec:results}).

\subsection{Impact handling}
A ground impact produces an acceleration that the model does not predict. If the estimator treated this acceleration as a real state change, the estimate would jump. We therefore gate the accelerometer: it is used only when $|\|\bm a\| - g| < 1.5$ m/s$^2$. When the gate is closed, the accelerometer row is dropped from the measurement set. In addition, the velocity process noise is multiplied by 40 during the impact window, so that the estimator can follow the change in velocity without a large penalty. The rule is heuristic, and it is tuned to the simulated impacts described in Section~\ref{sec:setup}.

\subsection{Sensor model for the baselines}
The complementary-filter baseline (LQR-CF) uses a PI complementary filter with integral gain $k_i = 0.3$ and a time constant of 1.0 s, and it uses the same accelerometer gating. It does not estimate the disturbance or the bias, which makes it the natural reference for an estimator without a model.

\section{Controller}
\label{sec:mpc}
\subsection{LQR baselines}
The LQR baselines use the continuous gain $\bm K(L)$ obtained from the algebraic Riccati equation, with weights $\bm Q_{\mathrm{LQR}} = \mathrm{diag}(30, 60, 400, 12, 4, 6)$ and $\bm R_{\mathrm{LQR}} = \mathrm{diag}(1.5, 0.6)$. Three variants are compared: LQR-oracle uses the ground-truth state, LQR-CF uses the complementary-filter estimate, and LQR-MHE uses the MHE estimate. A roll integral, a leaning-turn law, and a yaw controller are common to all variants.

\subsection{Offset-free constrained MPC}
The MPC uses the same linearized model in discrete time. It solves
\begin{align}
\min_{\bm u_{0:N-1}}\ & \sum_{k=0}^{N-1}\Big(\|\bm x_k - \bm x_{\mathrm{ref}}\|^2_{\bm Q} + \|\bm u_k - \bm u_s\|^2_{\bm R}\Big) \nonumber\\
\text{s.t.}\ & \bm x_{k+1} = \bm A_d\bm x_k + \bm B_d\bm u_k + \bm E_d\hat{\bm d},\quad |\bm u_k| \le \bm u_{\max},
\label{eq:mpc}
\end{align}
where $N = 40$ steps, $\bm u_s$ is the steady input, $\hat{\bm d}$ is the disturbance estimate from the MHE, and $\bm x_{\mathrm{ref}}$ is a steady target. The bounds $\bm u_{\max} = (16,\ 40)$\,N\,m apply to the total wheel torque and the hip torque. The wheel bound lies 4\,N\,m below the 20\,N\,m total of the two wheel actuators, and this margin is kept for yaw control. The hip bound lies 5\,N\,m below the 45\,N\,m actuator limit. The weights are $\bm Q = Q_{\mathrm{scale}}\,T_s\,\bm Q_{\mathrm{LQR}}$ with $Q_{\mathrm{scale}} = 2$, and $\bm R = T_s\,\bm R_{\mathrm{LQR}}$. The condensed quadratic program is solved with OSQP, warm-started from the previous solution.

Offset-free behavior comes from the disturbance estimate. Because $\hat{\bm d}$ enters the prediction, the controller holds the robot at the desired pose even when a payload or wind force is present. The steady target $\bm x_{\mathrm{ref}}$ is computed from $\hat{\bm d}$ and the commanded leg length, and it is low-pass filtered with a time constant of 0.5 s. Without this filter the target oscillates when the disturbance estimate changes.

To make the comparison fair, the discrete LQR gain used by the baselines is the unconstrained first-move gain of the MPC. We verified that the two gains agree to a relative error of $8\times10^{-10}$. Any difference between MPC and LQR in the results therefore comes from the constraints and the prediction horizon, and not from the weights.

\section{Simulation Setup}
\label{sec:setup}
The robot and environment are simulated in MuJoCo \cite{todorov2012}. The controller receives only simulated sensor data, with the sensor model in Table~\ref{tab:sensors}. The gyroscope bias is a constant offset that the estimator must learn. Encoder readings are quantized to 14 bits. The sensor stream has a two-sample transport delay, which the estimator compensates by predicting forward.

\begin{table}[t]
\centering
\caption{Simulated sensor model.}
\label{tab:sensors}
\begin{tabular}{@{}ll@{}}
\toprule
Sensor & Model \\
\midrule
Gyroscope noise & 0.01 rad/s (standard deviation) \\
Gyroscope bias & $(0.02,\ 0.04,\ 0.005)$ rad/s \\
Accelerometer noise & 0.15 m/s$^2$ (standard deviation) \\
Encoder resolution & 14 bits \\
Encoder velocity noise & 0.05 (standard deviation) \\
Transport delay & 2 samples, compensated by prediction \\
\bottomrule
\end{tabular}
\end{table}

Three scenarios are used.
\begin{itemize}
\item \textbf{A (driving, squatting, reversing, bowing).} A 12 s sequence of forward driving, a squat, a reverse, and a forward bow. This tests tracking and estimation under normal operation.
\item \textbf{B (unknown payload and wind).} A 2 kg payload is attached at $t = 2$ s and is not included in the controller model. A 10 N headwind is applied from 6 to 9 s, and a 0.6 m reposition is commanded from 9.5 to 11 s. The position error is reported separately during the payload phase and the wind phase.
\item \textbf{C (ball impact).} A 1.2 kg ball is thrown at the robot from the front at speeds from 3 to 14 m/s in steps of 0.5 m/s.
\end{itemize}
A grid over payloads of 0, 1.5, and 3 kg and headwinds of 0, 10, 15, and 20 N gives twelve combinations. Each combination is run with all four controllers. The reported metrics are the RMS speed error, the RMS pitch error and pitch-estimate error, the position error, and whether the robot fell.

\section{Results}
\label{sec:results}
\subsection{Scenario A: driving, squatting, reversing, bowing}
\begin{figure*}[!p!t]
	\centering
	\includegraphics[width=.8\textwidth]{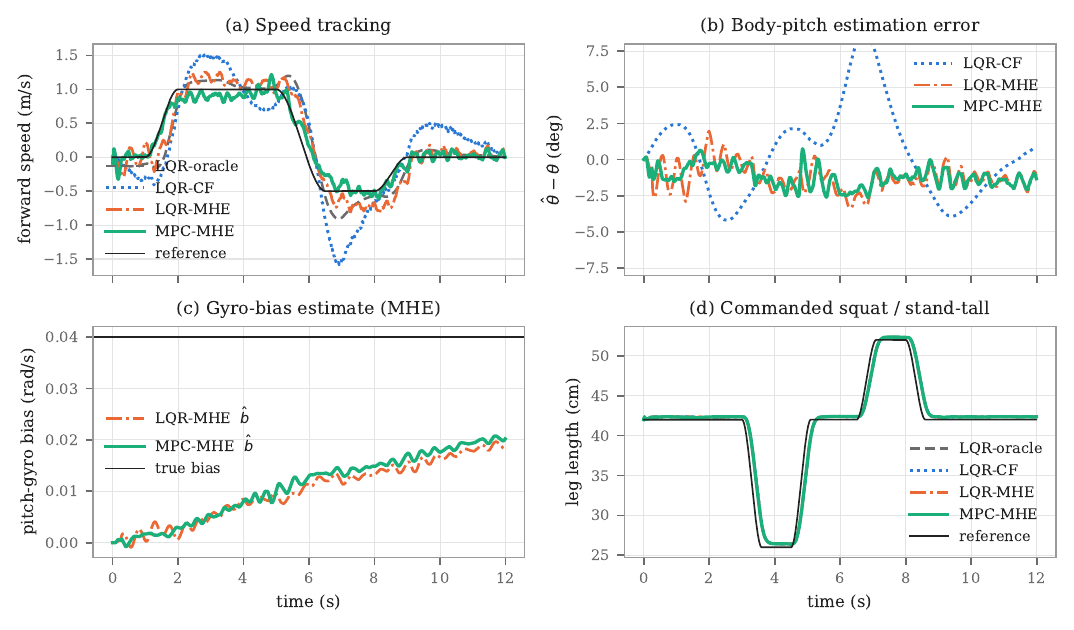}
	\caption{Scenario A. (a) Forward speed tracks the reference for the MHE-based controllers. (b) Body-pitch estimation error. (c) Gyroscope-bias estimate, which converges slowly toward its true value. (d) Commanded leg length during the squat.}
	\label{fig:A}
\end{figure*}
Table~\ref{tab:A} and Fig.~\ref{fig:A} summarize Scenario A. MPC-MHE had the lowest RMS speed error, at 0.114 m/s compared with 0.240 m/s for LQR-oracle. Its pitch-estimate error was 1.39$^\circ$, compared with 3.24$^\circ$ for LQR-CF, so the MHE reduced the pitch error of the complementary filter by more than half. LQR-MHE had a similar pitch-estimate error of 1.52$^\circ$. The MHE solve time had a mean of 3.5 ms and a maximum of 17.9 ms, and the maximum exceeds the 10 ms sample period. Section~\ref{sec:discussion} discusses this.

\begin{table}[t]
\centering
\caption{Scenario A results. Speed and pitch errors are RMS values over the 12 s run.}
\label{tab:A}
\begin{tabular}{@{}lcccc@{}}
\toprule
Controller & Speed err. (m/s) & Pitch err. ($^\circ$) & Max $|\theta|$ ($^\circ$) & Pitch-est. err. ($^\circ$) \\
\midrule
LQR-oracle & 0.240 & 1.20 & 14.2 & 0.00 \\
LQR-CF & 0.418 & 3.63 & 15.3 & 3.24 \\
LQR-MHE & 0.219 & 1.75 & 15.7 & 1.52 \\
MPC-MHE & \textbf{0.114} & 1.19 & 15.5 & 1.39 \\
\bottomrule
\end{tabular}
\end{table}

\subsection{Scenario B: unknown payload and wind}
\begin{figure*}[!p!t]
	\centering
	\includegraphics[width=.8\textwidth]{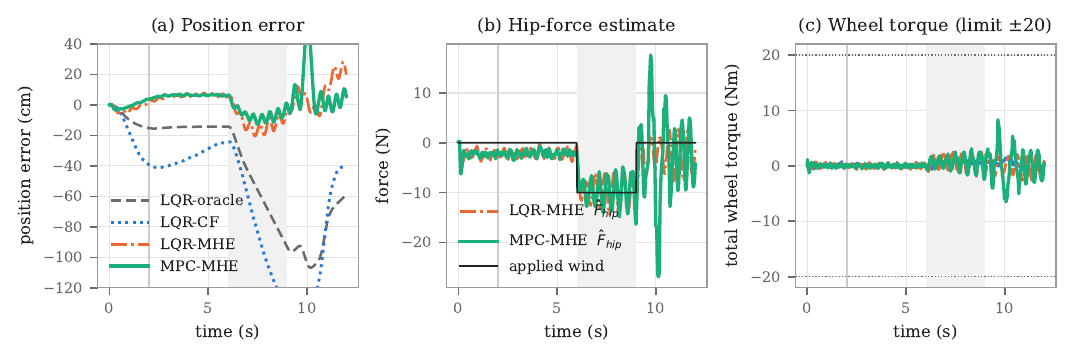}
	\caption{Scenario B. (a) Position error during the payload, headwind, and reposition phases. (b) Hip-force estimate for the two MHE controllers, with the applied headwind shown for reference. (c) Total wheel torque. The dotted lines mark the $\pm 20$ N$\cdot$m limit.}
	\label{fig:B}
\end{figure*}
Table~\ref{tab:B} and Fig.~\ref{fig:B} summarize Scenario B. With the ground-truth state, LQR-oracle drifted 81 cm in the headwind, because it had no disturbance estimate. LQR-CF drifted further, at 121 cm. Both MHE-based controllers kept the drift small. LQR-MHE drifted 10 cm in the wind and MPC-MHE drifted 6 cm. Both MHE controllers held the robot within about 6 to 7 cm during the payload phase.

Scenario B also exposes a cost of the MPC. Its RMS speed error was 0.53 m/s, compared with 0.32 m/s for LQR-MHE, and its RMS pitch error was 2.83$^\circ$, compared with 1.73$^\circ$. The increase comes mainly from the 0.6 m reposition, where the MPC response shows a small oscillation. The MPC also ends the run at 5.2 cm of final position error, compared with 22.6 cm for LQR-MHE, so the MPC settles more accurately despite the larger transient.

\begin{table}[t]
\centering
\caption{Scenario B results. Position errors are mean absolute values in the phase indicated.}
\label{tab:B}
\begin{tabular}{@{}lcccc@{}}
\toprule
Controller & Fell? & Payload (cm) & Wind (cm) & Final (cm) \\
\midrule
LQR-oracle & no & 14.3 & 80.9 & 61.9 \\
LQR-CF & no & 27.4 & 120.6 & 41.4 \\
LQR-MHE & no & 6.5 & 10.4 & 22.6 \\
MPC-MHE & no & 6.4 & \textbf{6.0} & \textbf{5.2} \\
\bottomrule
\end{tabular}
\end{table}

\subsection{Robustness grid}
Table~\ref{tab:grid} reports whether each controller survived each payload and headwind combination. LQR-MHE fell in three cells, all with a 3 kg payload, at 0, 15, and 20 N. MPC-MHE fell in one cell, at 3 kg and 20 N. Both controllers survived the other nine combinations.

The grid also shows that the MPC is not uniformly more accurate. At 0 kg and 20 N, the MPC-MHE mean wind error was 56 cm, compared with 9 cm for LQR-MHE. At 1.5 kg and 20 N, the MPC error was 53 cm, compared with 22 cm for LQR-MHE. Where both controllers survive, the advantage of the MPC is therefore in survival under heavy payloads, not in accuracy under strong wind. We return to this in Section~\ref{sec:discussion}.

\begin{table}[t]
\centering
\caption{Robustness grid. Entries give whether the robot survived and, if so, the mean position error in wind (cm).}
\label{tab:grid}
\begin{tabular}{@{}lcccc@{}}
\toprule
Payload / wind & LQR-oracle & LQR-CF & LQR-MHE & MPC-MHE \\
\midrule
0 kg / 0 N & ok & ok & ok & ok \\
0 kg / 10 N & ok, 111 & ok, 162 & ok, 3 & ok, 7 \\
0 kg / 15 N & ok, 167 & ok, 236 & ok, 6 & ok, 9 \\
0 kg / 20 N & ok, 221 & ok, 307 & ok, 9 & ok, 56 \\
1.5 kg / 0 N & ok & ok & ok & ok \\
1.5 kg / 10 N & ok, 87 & ok, 129 & ok, 9 & ok, 3 \\
1.5 kg / 15 N & ok, 133 & ok, 190 & ok, 15 & ok, 3 \\
1.5 kg / 20 N & ok, 179 & ok, 250 & ok, 22 & ok, 53 \\
3 kg / 0 N & ok & ok & \textbf{fell} & ok \\
3 kg / 10 N & ok, 68 & ok, 102 & ok, 22 & ok, 41 \\
3 kg / 15 N & ok, 106 & ok, 152 & \textbf{fell} & ok, 15 \\
3 kg / 20 N & ok, 144 & ok, 202 & \textbf{fell} & \textbf{fell} \\
\bottomrule
\end{tabular}
\end{table}

\subsection{Scenario C: ball impact}
\begin{figure*}[!p!t]
	\centering
	\includegraphics[width=.8\textwidth]{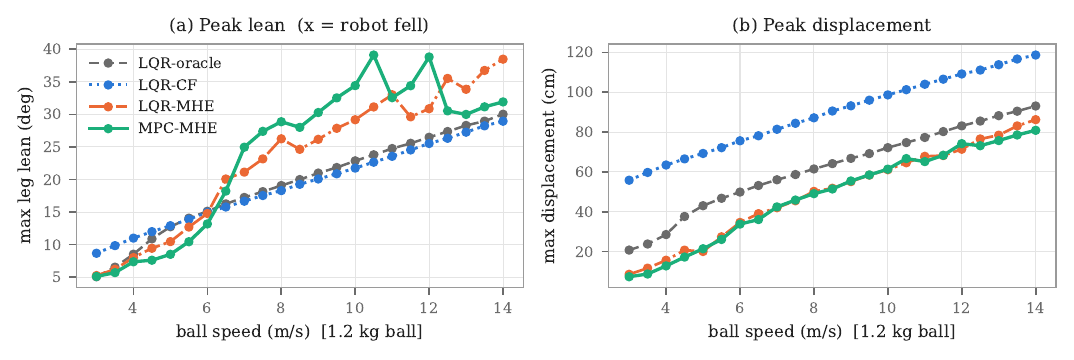}
	\caption{Scenario C. (a) Peak leg lean against ball speed; crosses mark failures. (b) Peak displacement against ball speed.}
	\label{fig:C}
\end{figure*}
All four controllers survived ball impacts up to the highest tested speed of 14 m/s. The differences appear in the peak lean. At 8 m/s, the peak lean was 19.1$^\circ$ for LQR-oracle, 18.3$^\circ$ for LQR-CF, 26.2$^\circ$ for LQR-MHE, and 28.9$^\circ$ for MPC-MHE (Fig.~\ref{fig:C}). The two estimator-based controllers lean more than the ground-truth controller. The impact-aware estimator is therefore not free: the velocity process noise that allows the estimate to follow an impact also makes the estimate less smooth. We have not isolated this effect, so the explanation should be treated as a conjecture.

\section{Discussion}
\label{sec:discussion}
The results support three observations. First, an estimator that models the disturbance removes most of the drift that a model-free filter leaves behind. In Scenario B, the LQR-CF drift in wind was 121 cm, compared with 10 cm for LQR-MHE. Second, the estimator change from complementary filter to MHE produced larger gains than the controller change from LQR to MPC in Scenario A, where the MPC reduced speed error by about half relative to LQR-MHE, and in Scenario B, where the MHE reduced wind drift by a factor of about twelve relative to LQR-CF. The MPC, by contrast, gave a larger lean in Scenario C. Third, the MPC improves survival under the heaviest payload, which is the regime where the input bounds are most likely to be active. We hypothesize that the MPC's constraint handling keeps the wheel torque inside the limit in these cases. The logs are consistent with this explanation, but they do not establish it.

The results also show disagreements that should be explained rather than hidden. In the 0 kg / 20 N cell, the MPC error is six times larger than that of LQR-MHE. This is consistent with the MPC trading accuracy for smoothness when the disturbance estimate is large. Similarly, the MPC pitch-estimate error in Scenario B is higher than that of LQR-MHE. These differences indicate that the MPC is not a uniform improvement, and that its benefit depends on the disturbance regime.

The limitations are substantial. All results are from simulation on flat ground, in the sagittal plane only. The payload is applied as an external force and is not added to the mass in the controller model. The estimated gyroscope bias converges slowly; after 12 s it reaches about half of the true value of 0.04 rad/s (Fig.~\ref{fig:A}c). Jumping and step-over maneuvers are not part of this evaluation. The MHE maximum solve time of 17.9 ms in Scenario A exceeds the 10 ms period, so a real-time implementation would need a shorter window, a warm start, or a faster solver.

\section{Conclusion}
\label{sec:conc}
A moving horizon estimator and an offset-free constrained MPC can keep a 9.95 kg wheel-legged robot balanced under an unknown payload and a headwind, in simulation with realistic sensing. The estimator provides the disturbance and bias estimates that the controller needs, and the MPC uses them under torque bounds. Across a 3 $\times$ 4 grid of payloads and headwinds, MPC-MHE fell in one case, compared with three for LQR with the same estimator. The advantage is concentrated in heavy payloads, and it comes with a cost in transient speed error and in lean during ball impacts.

The next stage is hardware validation. This requires the sensor model to be checked against measured IMU and encoder data, the impact rule to be tested on real ground impacts, and the MHE to be implemented within the real-time budget.

\end{document}